\documentclass[sigconf]{acmart}
\usepackage{graphicx}
\usepackage{multicol}
\usepackage{multirow}
\usepackage{epstopdf}
\usepackage{epsfig}
\usepackage{subfigure}
\usepackage{graphicx}
\usepackage{algorithm}
\usepackage{algorithmic}
\usepackage{setspace}
\usepackage{booktabs,bm} 
\acmDOI{10.475/123_4}

\acmISBN{123-4567-24-567/08/06}

\begin{document}
\title{Defensive Dropout for Hardening Deep Neural Networks under Adversarial Attacks}

\copyrightyear{2018} 
\acmYear{2018} 
\setcopyright{acmcopyright}
\acmConference[ICCAD '18]{IEEE/ACM INTERNATIONAL CONFERENCE ON COMPUTER-AIDED DESIGN}{November 5--8, 2018}{San Diego, CA, USA}
\acmBooktitle{IEEE/ACM INTERNATIONAL CONFERENCE ON COMPUTER-AIDED DESIGN (ICCAD '18), November 5--8, 2018, San Diego, CA, USA}
\acmPrice{15.00}
\acmDOI{10.1145/3240765.3264699}
\acmISBN{978-1-4503-5950-4/18/11}

\author{
Siyue Wang$^{1*}$, Xiao Wang$^{2*}$, Pu Zhao$^1$, Wujie Wen$^3$, David Kaeli$^1$, Peter Chin$^2$, Xue Lin$^1$
\\$^1$ Northeastern University \qquad\qquad\quad
$^2$ Boston university  \\
$^3$ Florida International University \quad\quad\
$^*$ Equal Contribution \\
\texttt{wang.siy@husky.neu.edu \quad kxw@bu.edu \quad zhao.pu@husky.neu.edu \quad wwen@fiu.edu \quad kaeli@ece.neu.edu \quad spchin@cs.bu.edu \quad xue.lin@northeastern.edu}
}

\begin{abstract}
Deep neural networks (DNNs) are known vulnerable to adversarial attacks. That is, adversarial examples, obtained by adding delicately crafted distortions onto original legal inputs, can mislead a DNN to classify them as any target labels. This work provides a solution to hardening DNNs under adversarial attacks through defensive dropout. Besides using dropout during training for the best test accuracy, we propose to use dropout also at test time to achieve strong defense effects. We consider the problem of building robust DNNs as an attacker-defender two-player game, where the attacker and the defender know each others' strategies and try to optimize their own strategies towards an equilibrium. Based on the observations of the effect of test dropout rate on test accuracy and attack success rate, we propose a defensive dropout algorithm to determine an optimal test dropout rate given the neural network model and the attacker's strategy for generating adversarial examples. We also investigate the mechanism behind the outstanding defense effects achieved by the proposed defensive dropout. Comparing with stochastic activation pruning (SAP), another defense method through introducing randomness into the DNN model, we find that our defensive dropout achieves much larger variances of the gradients, which is the key for the improved defense effects (much lower attack success rate). For example, our defensive dropout can reduce the attack success rate from 100\% to 13.89\% under the currently strongest attack i.e., C\&W attack on MNIST dataset.
\end{abstract}

\maketitle

\section{Introduction}

Deep neural networks (DNNs) are powerful models that achieve extraordinary performance in various speech and vision tasks, including speech recognition \cite{hinton2012deep}, natural language processing 
\cite{collobert2008unified}, 
scene understanding \cite{karpathy2015deep}, and object recognition \cite{krizhevsky2012imagenet, lecun1998gradient, lecun1989backpropagation}. 
However, recent studies \cite{szegedy2013intriguing,kurakin2016adversarial,goodfellow2014explaining} show that DNNs are vulnerable to adversarial attacks implemented by generating adversarial examples, i.e., adding imperceptible but well-designed distortions to the original legal inputs.
Delicately crafted adversarial examples can mislead a DNN to classify them as any target labels, while they appear recognizable and visually normal to human eyes.

Evidences have shown that audio/visual inputs sound/look like speeches/objects to DNNs but non-sense to humans \cite{carlini2016hidden,nguyen2015deep}.
Recently Kurakin, Goodfellow, and Bengio have demonstrated the existence of adversarial attacks not only in theoretical models but also the physical world \cite{kurakin2016adversarial}.
They mimicked the scenario of physical world application of DNNs by feeding the adversarial examples to a DNN through a cellphone camera to find that adversarial examples remain mis-classified by the DNN even when perceived through a camera.

Concerns have been aroused for applying DNNs in security-critical tasks.
The security properties of DNNs have been widely investigated from two aspects: (i) crafting adversarial examples to test the vulnerability of DNNs and (ii) enhancing the robustness of DNNs under adversarial attacks.
For the former aspect, adversarial examples have been generated by solving optimization problems \cite{szegedy2013intriguing,carlini2017towards,chen2017ead,athalye2018obfuscated,goodfellow2014explaining,papernot2016limitations}.
For the later aspect, research works have been conducted by either filtering out added distortions \cite{guo2017countering,bhagoji2017dimensionality,dziugaite2016study,xie2017mitigating} or revising DNN models \cite{papernot2016distillation,s.2018stochastic,feinman2017detecting} to defend against adversarial attacks. These two aspects mutually benefit each other towards hardening DNNs under adversarial attacks.

In this work, we provide a new solution to hardening DNNs under adversarial attacks through defensive dropout.
Dropout is a commonly used regulation method to deal with the overfitting issue due to limited training data \cite{srivastava2014dropout}.
As a regulation method, dropout is applied during training that for each training case in a mini-batch, a sub-network is sampled by dropping some of the units (i.e., neural nodes).
Based on some observations from preliminary experiments, we propose to use dropout also at test time as a defense method against adversarial attacks.
By introducing dropout to test time, we achieve shorter and fatter distributions of the gradients, which is the key for the improved defense effects (lower attack success rate) compared with another model-randomness-based defense method i.e., stochastic activation pruning (SAP) \cite{s.2018stochastic}.
For MNIST dataset, our defensive dropout reduces the attack success rate from 100\% to 13.89\% under the currently strongest attack i.e., C\&W attack \cite{carlini2017towards}, while the distillation as a defense \cite{papernot2016distillation} and the adversarial training \cite{tramer2017ensemble} are totally vulnerable under C\&W attack.
For CIFAR dataset, our defensive dropout reduces the attack success rate to 43.33\%, while SAP can only reduce the attack success rate to 77.78\% under C\&W attack based on the same neural network model.

The contributions of this work are summarized as following:

(i) We consider the problem of building robust DNNs as an attacker-defender two-player game, where the attacker and the defender know each others' strategies and try to optimize their own strategies towards an equilibrium.

(ii) We propose a defensive dropout algorithm that determines an optimal test dropout rate given the neural network model and the attacker's strategy for generating adversarial examples. Basically, we need to trade-off between the defense effects and test accuracy.

(iii) We explain the mechanism behind the outstanding defense effects by the proposed defensive dropout. The shorter and fatter gradient distributions make it difficult for the attacker to generate adversarial examples using the gradients from the sampled sub-networks.

\section{Related Work}

\subsection{Preliminaries}

In this paper we focus on neural networks used as image classifiers. 
In this case, the input images can be denoted as 3-dimensional tensors  $x\in \mathbb{R}^{h\times w\times c}$, where $h$, $w$ and $c$ denote the height, width and number of channels. 
For a gray scale image (e.g. MNIST), $c=1$; and for a colored RGB image (e.g. CIFAR-10), $c=3$. 
For both attacks and defends, all pixel values in the images are scaled to $[0,1]$ for easy calculation, and therefore a valid input image should be inside a unit cube in the high dimensional space.
We use model $F(x)=y$ to denote a neural network, where $F$ accepts an input $x$ and generates an output $y$. 

Suppose the neural network is an $m$-class classifier and the output layer performs softmax operation. Let $Z(x)$ denote the output of all layers except for the softmax layer, and we have $F({x}) = {\rm{softmax}}(Z({x})) = {y}$. The input to the softmax layer, $Z(x)$, is called logits. The element ${y_i}$ of the output vector ${y}$ represents the probability that input $x$ belongs to the $i$-th class. The output vector ${y}$ is treated as a probability distribution and its elements satisfy $0 \le {y_i} \le 1$ and $y_1+y_2+\dots+y_m=1$. The neural network classifies input $x$ according to the maximum probability i.e., $C(x) = \arg \mathop {\max }\limits_i {y_i}$. 

The adversarial attack can be either targeted or untargeted. Given an original legal input $x$ with its correct label $t^*$ and a target label $t \ne {t^*}$, the targeted adversarial attack is to find an input $x'$ such that $C(x') = t$ and $x$ and $x'$ are close according to some measure of the distortion between $x$ and $x'$. The input $x'$ is then called as an adversarial example. The untargeted adversarial attack is to find an input $x'$ satisfying $C(x') \ne t^*$ and $x$ and $x'$ are close according to some measure of the distortion. The untargeted adversarial attack does not specify any target label $t$ to mislead the classifier. In this work, we consider targeted adversarial attacks since they are believed stronger than untargeted attacks. 

The general problem of constructing adversarial examples can be formulated as:
\textbf{Given} an original legal input $x$, 
\begin{equation}
\begin{array}{l}
\text{minimize} \quad D(\delta) \\
\text{subject to}  \quad C(x+\delta) = t \\
\qquad \qquad \quad x+\delta \in [0,1]^n  
\end{array}
\end{equation}
where $\delta$ is the distortion added onto input $x$, $D(\delta)$ is a measure of the added distortion.

We need to measure the distortion between the original legal input $x$ and the adversarial example $x'=x+\delta$. ${L_p}$ norms are the most commonly used measures in the literature. The ${L_p}$ norm of the distortion is defined as:
\begin{equation}
{\left\| {x-x'} \right\|_p} = {\left( {\sum\limits_{i = 1}^n {{{\left| x_{i}-x'_{i} \right|}^p}} } \right)^{\frac{1}{p}}}
\end{equation}
We see the use of $L_0$, $L_1$, $L_2$, and $L_\infty$ norms in different attacks. 
\begin{itemize}
\item[-] \emph{${L_0}$ norm}: measures the number of mismatched elements between $x$ and $x'$. 
\item[-] \emph{${L_1}$ norm}: measures the sum of the absolute values of the differences between $x$ and $x'$.  
\item[-] \emph{${L_2}$ norm}: measures the standard Euclidean distance between $x$ and $x'$. 
\item[-] \emph{${L_\infty }$ norm}: measures the maximum difference between $x_i$ and $x'_{i}$ for all $i$'s.
\end{itemize}

\subsection{Attacks}
\subsubsection{Fault Injection \cite{liu2017fault}:} published in ICCAD 2017,  
proposes two kinds of fault injection attacks that only require slight changes to the DNN's
parameters to achieve misclassification: single bias attack (SBA) and gradient descent attack (GDA). SAB is able to achieve misclassification by modifying only one bias value in the network. And GDA achieves higher stealthiness and efficiency by using layer-wise searching and modification compression techniques.
It implements very efficient attacks on MNIST and CIFAR-10 datasets.
This work perceives the DNN attack problem from a different angle, i.e., modifying the DNN models, while all the other attacks and defends mentioned in this paper assume the modifications are performed onto the inputs.

\subsubsection{Fast Gradient Sign Method (FGSM) \cite{goodfellow2014explaining}:} is an $L_\infty$ attack and utilizes the gradient of the loss function to determine the direction to modify the pixels. They are designed to be fast, rather than optimal. They can be used for adversarial training by directly changing the loss function instead of explicitly injecting adversarial examples into the training data. FGSM generates adversarial examples following:
\begin{equation}
x' = x - \epsilon \cdot \text{sign} (\nabla(loss_{F,t}(x)))
\end{equation}
where $\epsilon$ is the magnitude of the added distortion, $t$ is the target label.
Using backpropagation, FGSM calculates the gradient of the loss function with respect to the label $t$ to determine the direction to change the pixel values. 

\subsubsection{Jacobian-based Saliency Map Attack (JSMA) \cite{papernot2016limitations}:} is an $L_0$ attack and uses a greedy algorithm that picks the most influential pixels by calculating Jacobian-based Saliency Map and modifies the pixels iteratively. The computational complexity of this attack method is very high.

\subsubsection{C\&W \cite{carlini2017towards}:} is a series of $L_0$, $L_2$, and $L_\infty$ attacks that achieve 100\% attack success rate with much lower distortions comparing with the above-mentioned attacks. In particular, the C\&W $L_2$ attack is superior to other $L_2$ attacks because it uses a better objective function.
C\&W formulates the problem of generating adversarial examples in an alternative way that can be better optimized:
\begin{equation}
\begin{array}{l}
\text{minimize} \quad D(\delta) + c\cdot f (x+\delta) \\
\text{subject to} \quad  x+\delta \in [0,1]^n \\
\end{array}
\end{equation}
where $c>0$ is a constant to be chosen and objective function $f$ has the following form:
\begin{equation}
  f(x+\delta) = \text{max} (\text{ max}\{Z(x+\delta)_i : i \neq t\} - Z(x+\delta)_t, -\kappa)   
\end{equation}
Here, $\kappa$ is a parameter that controls the confidence in attacks.
Stochastic gradient decent methods can be used to solve this problem. 
For example, the Adam optimizer \cite{KingmaB2015adam} is used due to its fast and robust convergence behavior.

\subsection{Defense Methods}
\subsubsection{Adversarial Training \cite{tram2018ensemble}:} injects adversarial examples with correct labels into the training dataset and then retrains the neural network, thus increasing robustness of DNNs under adversarial attacks.

\subsubsection{Distillation as a Defense \cite{papernot2016distillation}:}
introduces \emph{temperature} $T$ into the softmax layer and uses a higher temperature for training and a lower temperature for testing. The training phase first trains a teacher model that can produce soft labels for the training dataset and then trains a distilled model using the training dataset with soft labels. The distilled model with reduced temperature will be preserved for testing. The modified softmax function utilized in the distilled model is given by:

\begin{equation}
 \text{softmax} (x, T)_i = \frac{e^{{Z(x)_i}/T}}{\sum_j  e^{{Z(x)_j}/T}}
\end{equation}
where $Z(x)_i$ is the $i$-th logit corresponding to the last hidden layer output before softmax.

\subsubsection{Stochastic Activation Pruning (SAP) \cite{s.2018stochastic}:} 
SAP uses $L_1$ normalization to calculate the multinomial distribution regarding every activation layer.
For every hidden activation vector $h^{i} \in \mathbb{R}^{a^{i}}$ at the $i$-th layer, SAP defines the probability $p^{i}_{j}$ of whether or not to prune the $j$-th activation output $h^{i}_{j}$ with the following, 
\begin{equation}
 p^{i}_{j} = \frac{|h^{i}_{j}|}{\sum_{k=1}^{a^{i}} |h^{i}_{k}|}
\end{equation}

The remaining activation outputs are scaled up according to the pruning probability and the number of outputs kept at this layer by using the reweighting factor,
\begin{equation}
 q^{i}_{j} = 1 - (1 - p^{i}_{j})^{r^i_p}
\end{equation}
where $r^i_p$ is the number of outputs sampled.

\section{Proposed Defensive Dropout}
\subsection{Dropout Preliminaries}
\begin{figure}
\centering                                        
\includegraphics[width=0.48\textwidth]{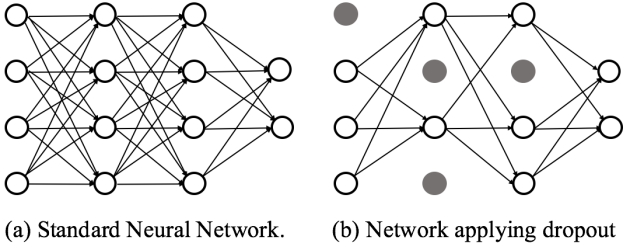} 
\caption{(a) A standard neural network with 2 hidden layers. (b) A sub-network produced by applying dropout. Units in grey color are dropped from the whole network.}\label{Fig:DropoutModel}
\end{figure}

Motivated by the mixability theory in the evolutionary biology \cite{livnat2010sex}, dropout was proposed as a regularization method in machine learning to prevent the overfitting issue with limited training data \cite{srivastava2014dropout}.
The term ``dropout'' refers to dropping out units (hidden and visible) in a neural network.
By dropping a unit out, the unit along with all its incoming and outgoing connections are temporarily removed from the network.
Fig. \ref{Fig:DropoutModel} shows applying dropout to a neural network amounts to sampling a sub-network from it.

The feedforward operation of a standard neural network can be described as:
\begin{align}
z^{(l+1)}_i & = \mathbf{w}^{(l+1)}_i \mathbf{y}^{(l)}+b^{(l+1)}_i  \\
y^{(l+1)}_i & = f(z^{(l+1)}_i)
\end{align}
where $\mathbf{y}^{(l)}$ denotes the vector of outputs from layer $l$, $\mathbf{z}^{(l)}$ denotes the vector of inputs to layer $l$, $\mathbf{W}^{(l)}$ and $\mathbf{b}^{(l)}$ are weight matrix and bias vector of layer $l$, and $f$ is any activation function. With dropout, the feedforward operation becomes \cite{hinton2012improving}:
\begin{align}
r^{(l)}_j & \sim \text{Bernoulli}(p) \\
\widetilde{\mathbf{y}}^{(l)} & = \mathbf{r}^{(l)}\odot \mathbf{y}^{(l)} \\
z^{(l+1)}_i & = \mathbf{w}^{(l+1)}_i \widetilde{\mathbf{y}}^{(l)}+b^{(l+1)}_i  \\
y^{(l+1)}_i & = f(z^{(l+1)}_i)
\end{align}
where $\mathbf{r}^{(l)}$ is a vector of independent Bernoulli random variables, each with probability $p$ of being 1, and $\odot$ denotes an element-wise product.

A neural network with $n$ units can been seen as a collection of $2^n$ sampled sub-networks, which all share weights so that the total number of parameters is still $O(n^2)$.
During training a neural network with dropout, stochastic gradient descent is used as in standard training, except that for each training case in a mini-batch, a sub-network is sampled by dropping out units, and forward and backpropagation for that training case are done only on this sub-network.
The gradients for each parameter are averaged over the training cases in each mini-batch.
Therefore, training a neural network with $n$ units using dropout can be seen as training a collection of $2^n$ sub-networks with extensive weight sharing.

\begin{figure}
\centering                                        
\includegraphics[width=0.45\textwidth]{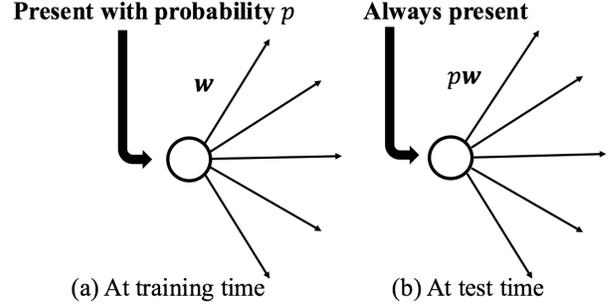} 
\caption{(a) At training time, a unit presents with probability $p$. (b) At test time, the unit is always present and the outgoing weights are multiplied by $p$.}\label{Fig:WeightScale}
\end{figure}

The purpose of applying dropout is to prevent units from co-adapting too much by combining the predictions of many sub-networks with shared weights.
However, at test time, it is not feasible to explicitly average the predictions from exponentially many sub-networks.
A very simple approximate averaging method is to use a single neural network at test time without dropout, the weights of which are scaled-down versions of the trained weights.
If a unit presents with probability $p$ during training with dropout, the outgoing weights of that unit are multiplied by $p$ at test time, as shown in Fig. \ref{Fig:WeightScale}.
This ensures that the output at test time is the same as the expected output at training time.

\subsection{Defensive Dropout Implementations in Training and Test}
Dropout is a commonly used regularization method. 
To achieve very good test accuracy, in practice dropout is usually applied to units in the fully-connected layer close to the output layer of the neural network \cite{goodfellow2016deep}.
Also, the dropout rate $r=1-p$, instead of the probability $p$ of presence for a unit is used during training with dropout \cite{bouthillier2015dropout}.
For each training case in a mini-batch, the units are dropped with rate $r$ and a sub-network is sampled for the training case. 
The gradient for each parameter (weight) is then calculated based on the sampled sub-network. 
Please note that, for a unit with rate $r$ of being dropped, if it presents in the sub-network, we need to divide the output of its activation function by $1-r$ when evaluating the loss function for gradient calculation. 
This is for making the output at test time roughly the same as the expected output at training time.  
If a parameter is not used in the sub-network, a zero gradient is set for that.
Gradients for each parameter are averaged over all training cases in the mini-batch.
When dropout is applied as a regulation method to deal with overfitting, at test time the whole neural network without dropout is used, but with the output of the activation function divided by $1-r$ if the unit is dropped with rate $r$ during training.

Intuitively, introducing randomness into the test time can also help to harden deep neural networks against adversarial attacks.
Therefore, we propose to apply dropout also \textbf{at the test time} as a defense method.
If dropout was applied to units in a specific layer during training with dropout rate $r$, we are going to apply dropout to the same layer at test time with dropout rate $r'$.
For each test case, units are dropped with rate $r'$ and a sub-network is sampled for it.
To have roughly the same expected output as the whole neural network at test time, we also need to scale-up the activation functions of the retained units in the dropout layer of the sub-network by $\frac{1}{1-r'}$.
Please note that $r$ is optimized during deep neural network training, and $r'$ is the optimization variable of our defense method to be derived by the Algorithm in Section \ref{sec:algorithm}.

Our work aims at defending against the strongest attacks, i.e., the \emph{white-box} attacks, that is, the attacker has perfect information about the neural network architecture and parameters.
Therefore, when we defend against adversarial attacks, we assume the attacker knows not only the complete neural network model but also the stochastics in the model (i.e., which layer applied dropout and the dropout rates $r$ and $r'$).
Specifically, when the attacker generating adversarial examples by solving an optimization problem based on stochastic gradient descent, the gradients are calculated in the similar manner as training with dropout i.e., using sampled sub-networks and the activation functions scaled-up by $\frac{1}{1-r'}$.
By doing this, we give the attacker full access to the neural network model and we are able to evaluate our defense method against the strongest white-box attacks.

\subsection{Observations and Motivations}

\begin{figure}[htbp]
\centering                    
\begin{minipage}[t]{0.45\textwidth}
\subfigure[Test Accuracy]{
\centering                                  \includegraphics[width=1\textwidth]{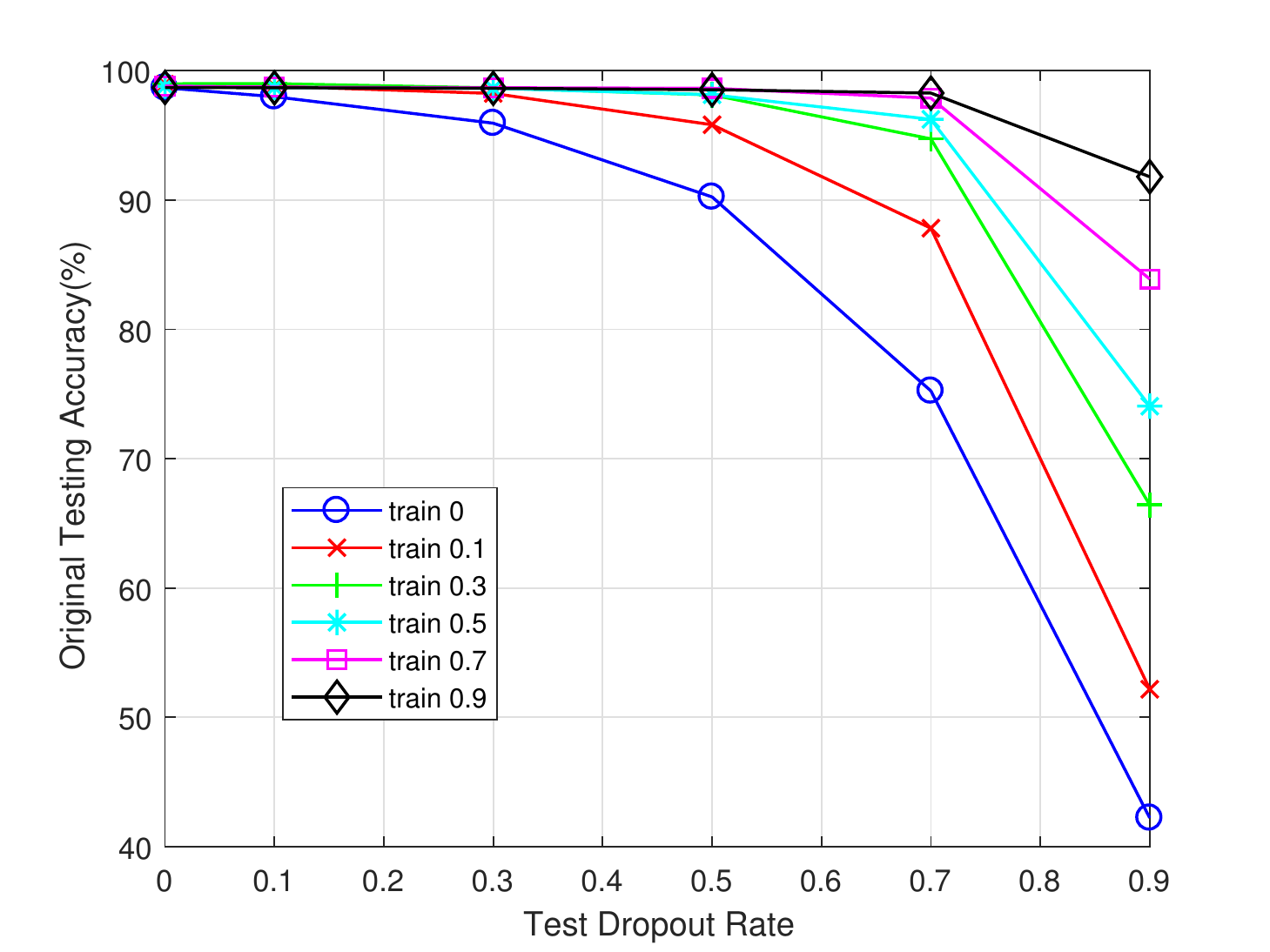}}
\subfigure[C\&W Attack Success Rate]{ 
\centering                                  \includegraphics[width=1\textwidth]{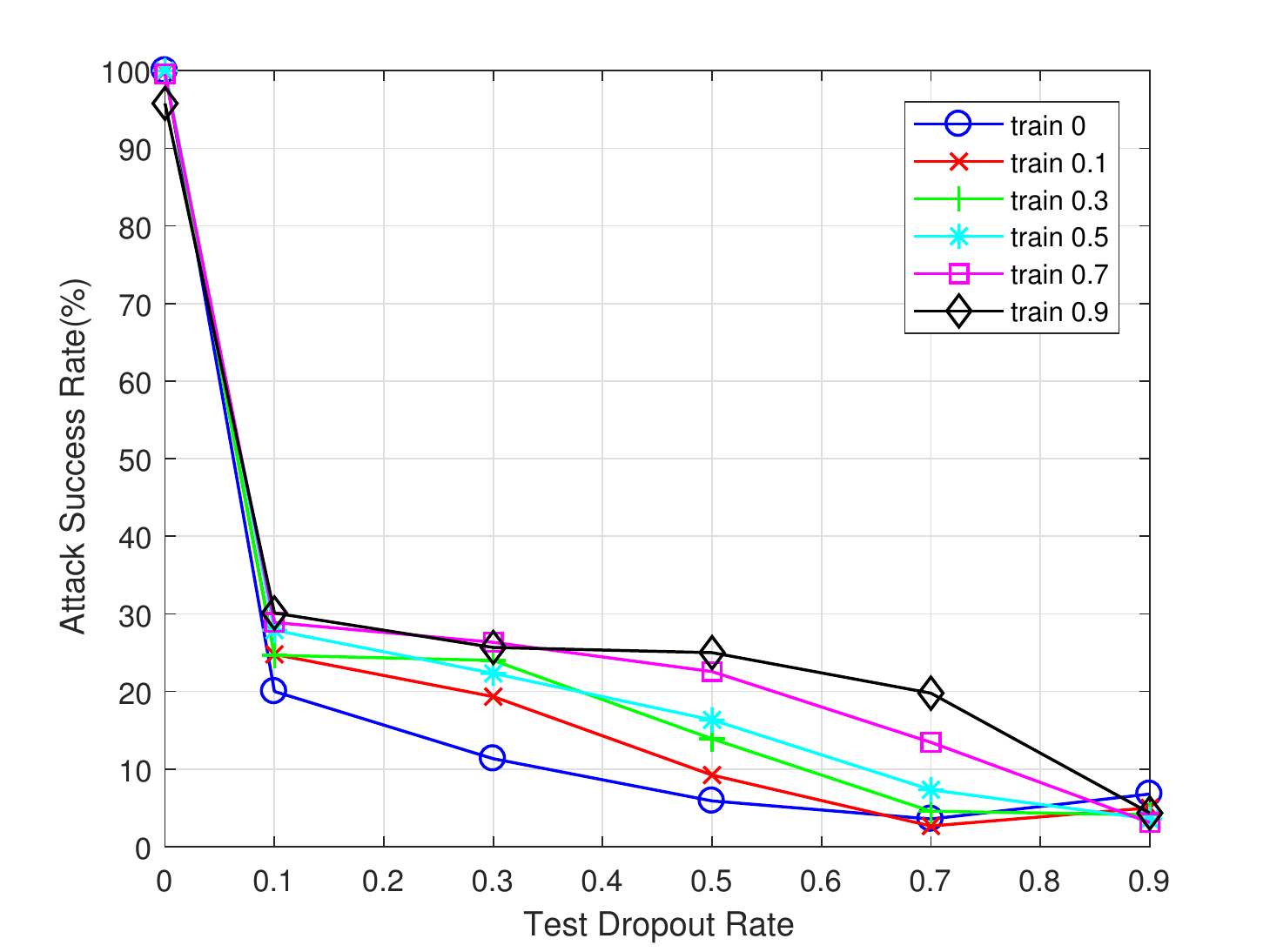}}
\subfigure[C\&W Attack $L_2$ Norm]{ 
\centering              
\includegraphics[width=1\textwidth]{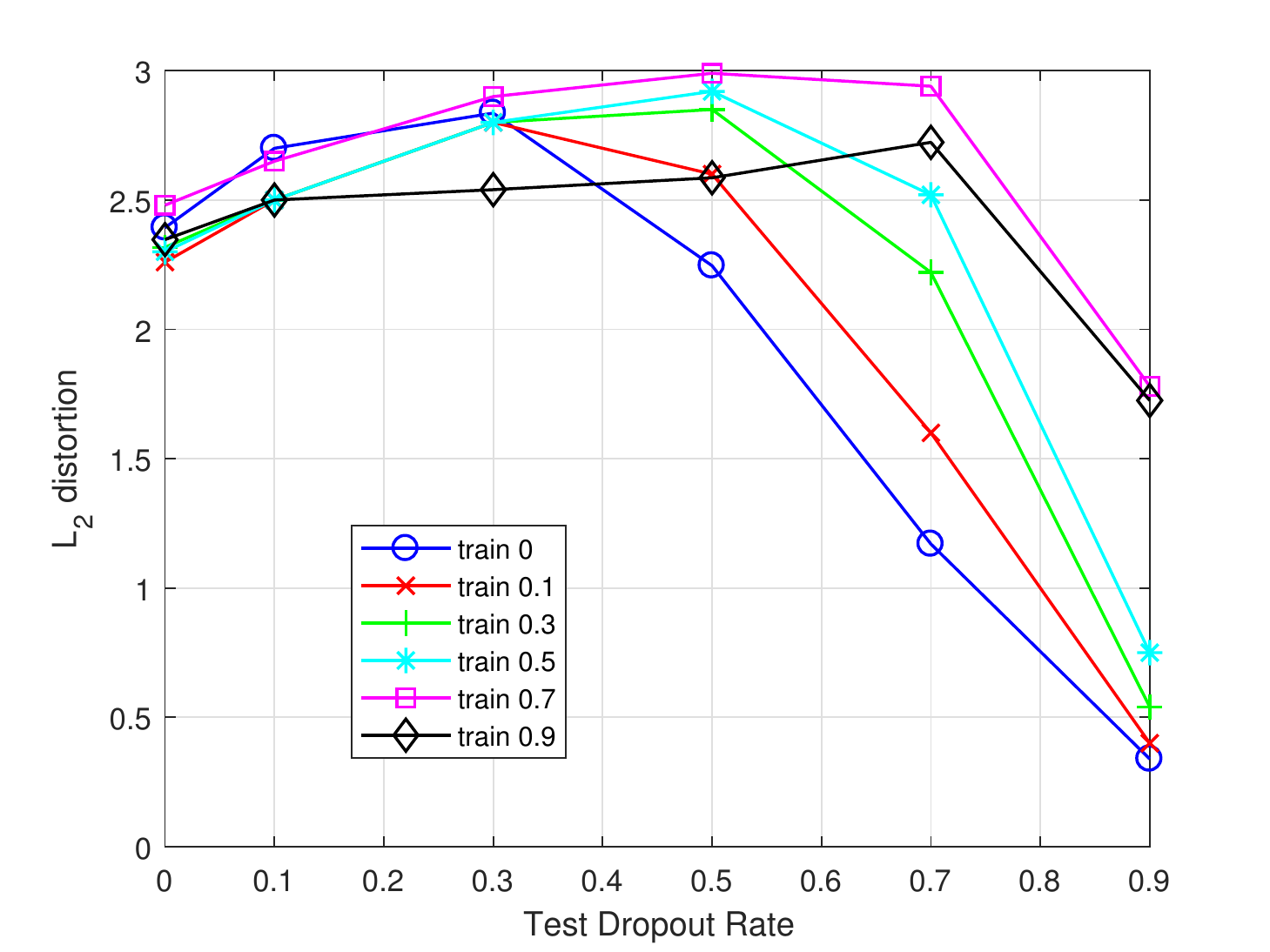}}
\end{minipage}
\caption{(a) Test accuracy, (b) Attack success rate, and (c) $L_2$ norm under C\&W attack on MNIST dataset using different training dropout rates and test dropout rates.}\label{Fig:C&WMNIST} 
\end{figure}

We perform some preliminary experiments that motivate and support our defensive dropout method.
We pick the currently strongest attack, i.e., C\&W attack \cite{carlini2017towards} and experiment on different training dropout rates and different test dropout rates to analyze test accuracy and defense effects.
Fig. \ref{Fig:C&WMNIST} presents the results, where the x-axes denote test dropout rate and each curve represents one training dropout rate.
From Fig. \ref{Fig:C&WMNIST} (a), we can observe that test accuracy decreases with increasing test dropout rate.
Also, a training dropout rate of 0.3 achieves the highest test accuracy for MNIST dataset. The increase in test accuracy is more prominent in other datasets. 
For example, a training dropout rate of 0.7 increases the test accuracy by 7.5\% in our neural network model on CIFAR-10 dataset.
In summary, the decrease in test accuracy due to the test dropout rate can be compensated in some extent by the training dropout rate.

Fig. \ref{Fig:C&WMNIST} (b) and (c) demonstrate the defense effects of training and test dropout rates.
In general, increasing test dropout rate can reduce the attack success rate, see Fig. \ref{Fig:C&WMNIST} (b). 
And the $L_2$ norm of the added distortions in the adversarial examples reaches a peak at certain test dropout rate, see Fig. \ref{Fig:C&WMNIST} (c).
The solution for defending against C\&W attack on MNIST dataset is to use a test dropout rate of 0.5 when the training dropout rate is 0.3, which loses 0.8\% test accuracy but decreases the attack success rate from 100\% to 13.89\% with the largest $L_2$ norm of the distortion (the peak point in Fig. \ref{Fig:C&WMNIST} (c)), indicating that the added distortion in the adversarial examples might be large enough to be recognized by humans.
Please note that under C\&W attack, the distillation as a defense \cite{papernot2016distillation} and the adversarial training \cite{tramer2017ensemble} could not decrease the attack success rate at all \cite{carlini2017towards}, i.e., still 100\% attack success rate under these defense methods.

\begin{figure*}
\centering                                        
\includegraphics[scale=0.52]{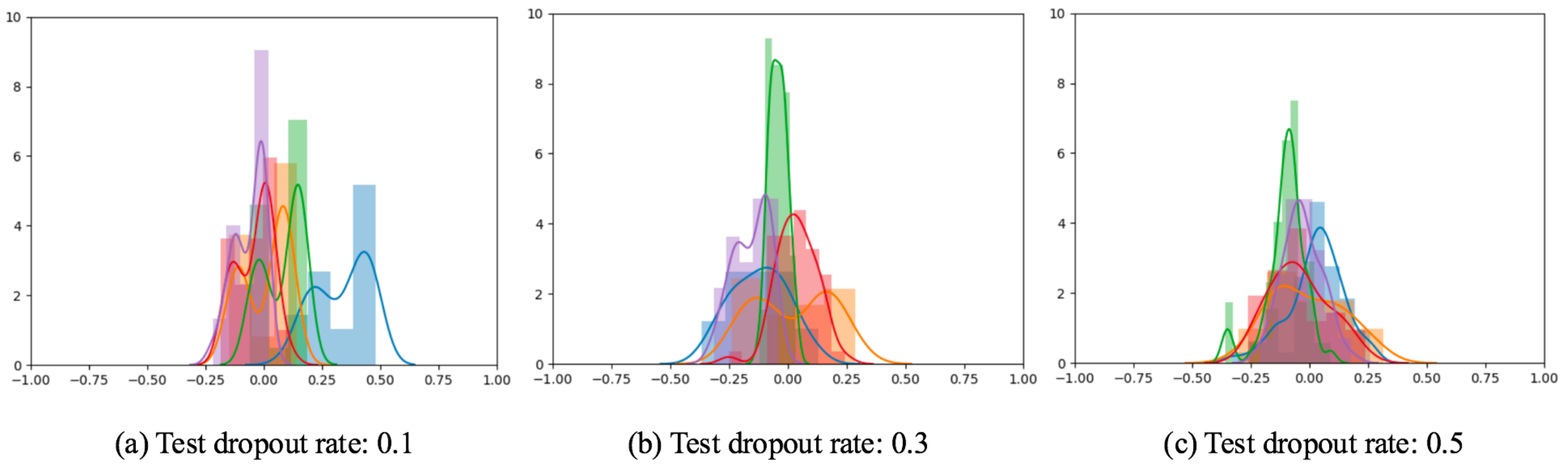} 
\centering
\includegraphics[scale=0.52]{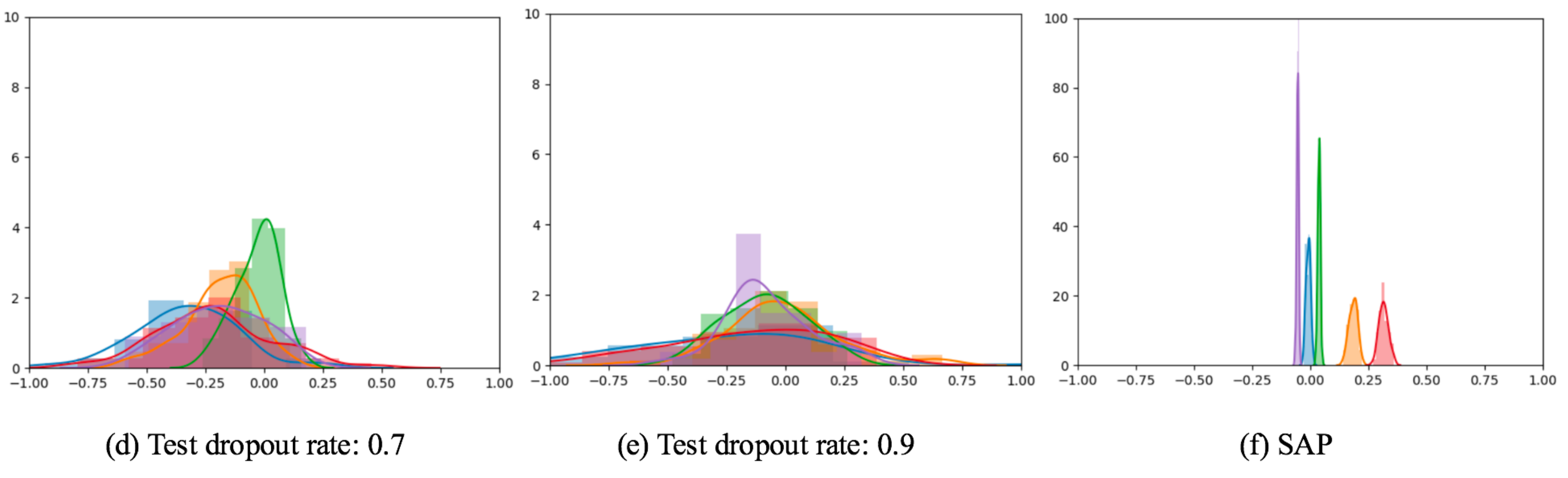}
\caption{Probability density of sampled gradients when generating adversarial example using C\&W attack on CIFAR-10.
The same neural network architecture, the same original input image, and the same training dropout rate of 0.7 for the best test accuracy is used throughout (a)$\sim$(f).
Histogram and the corresponding fitted probability density curve in each color denote one out of $32\times32\times3$ dimensions in the sampled gradients and include 50 data points. (a)$\sim$(e) are from our proposed defensive dropout for different test dropout rate, and (f) is from stochastic activation pruning (SAP).}\label{Fig:Gradients}
\end{figure*}

We also investigate the mechanism behind the outstanding defense effects achieved by the proposed defensive dropout.
It is intuitive to explain the defense effects as model randomness through adding dropout at test time.
However, there are also other defenses through introducing model randomness, such as stochastic activation pruning (SAP) \cite{s.2018stochastic} and migration through randomization (MTR) \cite{xie2017mitigating}, that only achieve limited defense effects against C\&W.
We are trying to explain the mechanism in a different way.
Fig. \ref{Fig:Gradients} is plotted for the probability density of uniformly sampled gradients when generating adversarial example using C\&W attack on CIFAR-10 dataset. 
Please note that the gradients have $32\times32\times3$ dimensions, and therefore we select 5 dimensions for visualization, each presented by a color and each containing 50 data points throughout Fig. \ref{Fig:Gradients} (a)$\sim$(f).
Also for fair comparison, we use the same neural network model with training dropout rate of 0.7 for the best test accuracy, and the same original input image for Fig. \ref{Fig:Gradients} (a)$\sim$(f), including our defensive dropout and SAP.

From Fig. \ref{Fig:Gradients} (a)$\sim$(e), with increasing test dropout rate, the probability densities become shorter and fatter, demonstrating increasing variances of the gradients, which is the key for the improved defense effects (decreasing attack success rate) with increasing test dropout rate.
The larger variances of the gradients, the more difficult for the attacker to generate effective adversarial examples by using stochastic gradient descent when solving the optimization problem.
It cross-validates the conclusion from Fig. \ref{Fig:C&WMNIST} (a) and (b).
Of course, we could not use the largest test dropout rate for the strongest defense effects, because the test accuracy might be very low.
We need to trade-off defense effects for the test accuracy.
Fig. \ref{Fig:Gradients} (f) is the probability densities of the gradients from stochastic activation pruning (SAP) \cite{s.2018stochastic}, which shows very small variances of the gradients comparing with our defense method.
That is the reason our defensive dropout outperforms SAP.

\subsection{Defensive Dropout Algorithm}\label{sec:algorithm}

Towards hardening deep neural networks under adversarial attacks, the attacker and the defender improve their own strategies like in a two-player game.
In such a game, the attacker and the defender know each others' strategies and try to optimize their own strategies towards an equilibrium.
The defender can benefit from the improvement of the attacker's strategy.
Therefore, we need to take into consideration the attacker's strategy of generating adversarial examples when designing our defense.

Based on the observations on the test dropout rate's effect on test accuracy and attack success rate, we design the defensive dropout algorithm that helps us to determine an optimal test dropout rate given the neural network model and the attacker's strategy for generating adversarial examples.
In the algorithm, we also optimize the training dropout rate along with the test dropout rate, but that is only for the purpose of training neural network for the best test accuracy.
Basically, we first train a neural network model $F$ by finding a proper training dropout rate $r$.
Then we fix the model $F$ and search for the largest test dropout rate $r'$ for the strongest defense effects while satisfying the test accuracy requirement.
Pseudo code of the defensive dropout algorithm is given in Algorithm \ref{ALG:DEFENSE}.

\begin{algorithm}[h]
\caption{Defensive Dropout Algorithm}
\label{ALG:DEFENSE}
\begin{algorithmic}[1]
\REQUIRE 
$\mathbf{X}$: Dataset \\
$F$: Neural Network Model \\
$\mathbf{\pi}$: Attacker Strategy (e.g., C\&W, FGSM, JSMA) \\
$\varepsilon$: Maximum decrease of test accuracy \\
\ENSURE 
$r'$: Test Dropout Rate \\
$r$: Training Dropout Rate \\

\STATE Retrain neural network model $F$ using a proper training dropout rate $r\in [0.3, 0.7]$ and obtain the best test accuracy $a_{test}^{max}$;
\STATE $a\leftarrow a_{test}^{max}$;
\STATE $r'\leftarrow 0$;
\WHILE{$a>a_{test}^{max}-\varepsilon$}
\STATE $r' \leftarrow r'+$ step\_size;
\STATE Generate adversarial examples using attacker strategy $\pi$, neural network model $F$, and test dropout rate $r'$;
\STATE Evaluate test accuracy $a$ using neural network model $F$ and test dropout rate $r'$;
\STATE Evaluate attack success rate using neural network model $F$ and test dropout rate $r'$;
\ENDWHILE
\end{algorithmic}
\end{algorithm}

\section{EXPERIMENTAL RESULTS}

\subsection{Setup}

As in other attack and defense work, we are also using the two datasets: \emph{MNIST} and \emph{CIFAR-10}.
The \emph{MNIST} dataset (Modified National Institute of Standards and Technology database) \cite{yann1998mnist} is a collection of handwritten digits that is commonly used for training and test various machine learning tasks. It consists of  $70,000$ $  28\times28$ ($60,000$ training images and $10,000$ test images) grey-scale images for digits $0$ to $9$. The \emph{CIFAR-10} dataset (Canadian Institute For Advanced Research) \cite{krizhevsky2009learning} is a collection of color images. It contains $60,000$ $32\times32$ images in $10$ different classes (e.g., cars, birds, airplanes, etc.). 


For DNN models in our experiment,  we use standard convolutional neural networks with 4 convolutional layers and 2 fully-connected layers. This architecture has been used as standard model in many previous work \cite{carlini2017towards, papernot2016distillation}. 
While the overall neural network architecture for MNIST or CIFAR-10 dataset is the same, the size of the neural network model for CIFAR-10 is slightly larger than that for MNIST, since CIFAR-10 images have higher resolution.
The activation function is rectified linear unit (ReLU) for all convolutional and fully connected layers. The architectures of the neural network models for MNIST and CIFAR-10 are summarized in Table  \ref{table_model_architecture}.
In both models we apply defensive dropout to \emph{Fully connected layer 1}.
After training, the models achieve the state-of-the-art 99.4\% and 80\% test accuracies for \emph{MNIST} dataset and \emph{CIFAR-10} dataset, respectively.

\begin{table}[h]
\caption{Architectures of neural network models for MNIST and CIFAR-10}
\label{table_model_architecture}
\begin{center}
\scalebox{0.9}{
\begin{tabular}{|c|c|c|}
    \hline
     & Model for MNIST  &  Model for CIFAR-10\\
    \hline
     Conv layer & 32 filters with size (3,3) & 64 filters with size (3,3) \\
     Conv layer & 32 filters with size (3,3) & 64 filters with size (3,3) \\
     Pooling layer & pool size (2,2) & pool size (2,2) \\
     Conv layer & 64 filters with size (3,3) & 128 filters with size (3,3) \\
     Conv layer & 64 filters with size (3,3) & 128 filters with size (3,3) \\
     Pooling layer & pool size (2,2) & pool size (2,2) \\
     Fully connected 1 & 200 units & 256 units \\
     Fully connected 2 & 200 units & 256 units \\
     Output layer & 10 units & 10 units \\
     \hline
\end{tabular}}
\end{center}
\end{table}

We implement FGSM, JSMA, and C\&W attacks based on the CleverHans package \cite{papernot2016cleverhans}.
For FGSM, we use a fixed $\epsilon = 0.25$ as suggested in the original paper \cite{goodfellow2014explaining}. 
For JSMA, we use the codes from CleverHans directly.
For C\&W, we perform $10$ iterations of binary search for constant $c$. With a selected $c$, we then run $100$ iterations of gradient decent with the Adam optimizer.
We compare with other defenses such as adversarial training \cite{tram2018ensemble}, distillation as a defense \cite{papernot2016distillation}, and stochastic activation pruning (SAP) \cite{s.2018stochastic}, among which we implement SAP ourselves, and defense effects of the adversarial training and distillation as a defense are cited from \cite{carlini2017towards}. 
\subsection{Results}

We use two metrics to evaluate the defense effects against adversartial attacks, i.e., attack success rate (ASR) and $L_2$ norm of the distortion. The lower attack sucssess rate and the higher $L_2$ norm imply the stronger defense effects.

\begin{figure}[htbp]
\centering                    
\begin{minipage}[t]{0.45\textwidth}
\subfigure[Test Accuracy]{
\centering                                  \includegraphics[width=1\textwidth]{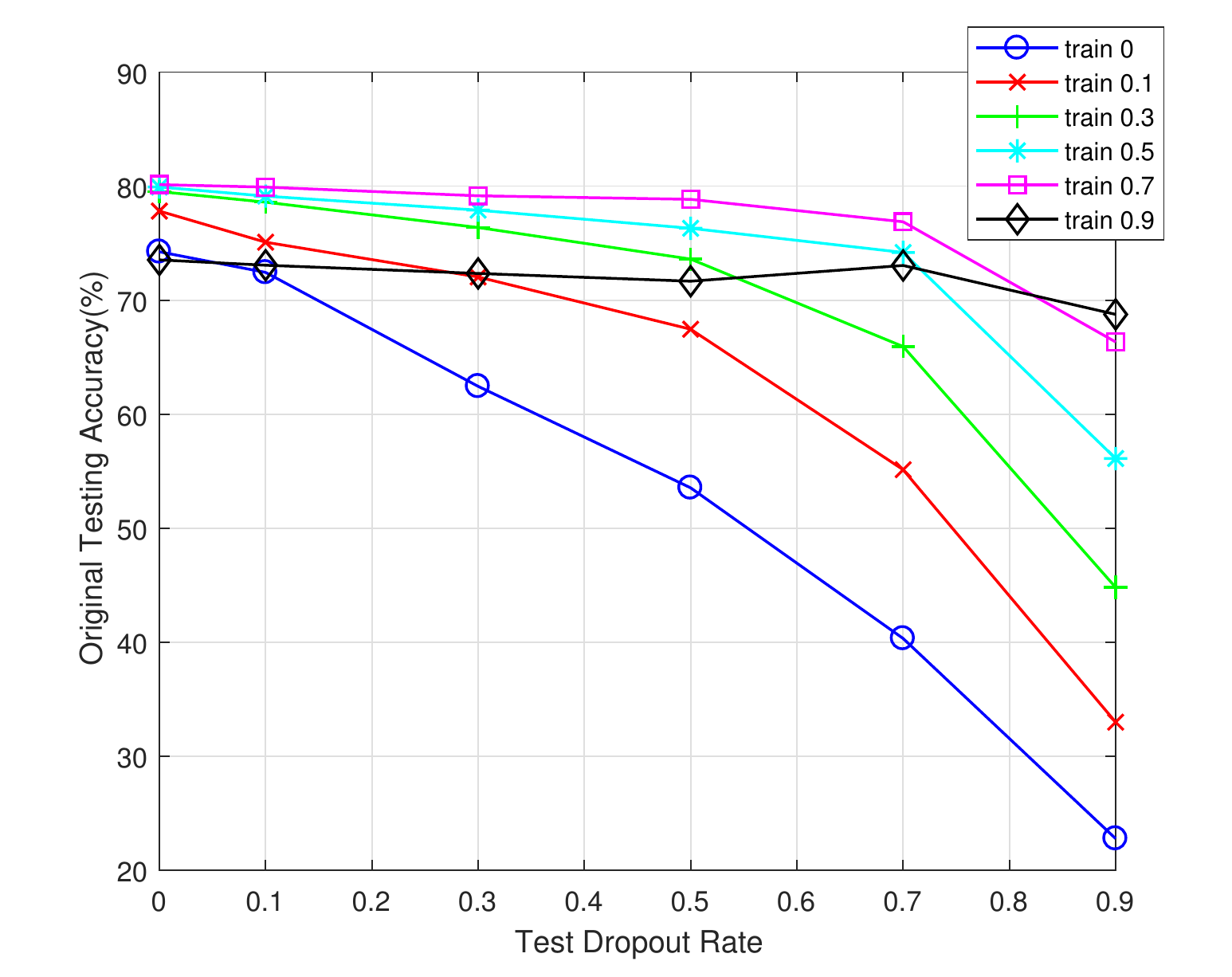}}
\subfigure[C\&W Attack Success Rate]{ 
\centering                                  \includegraphics[width=1\textwidth]{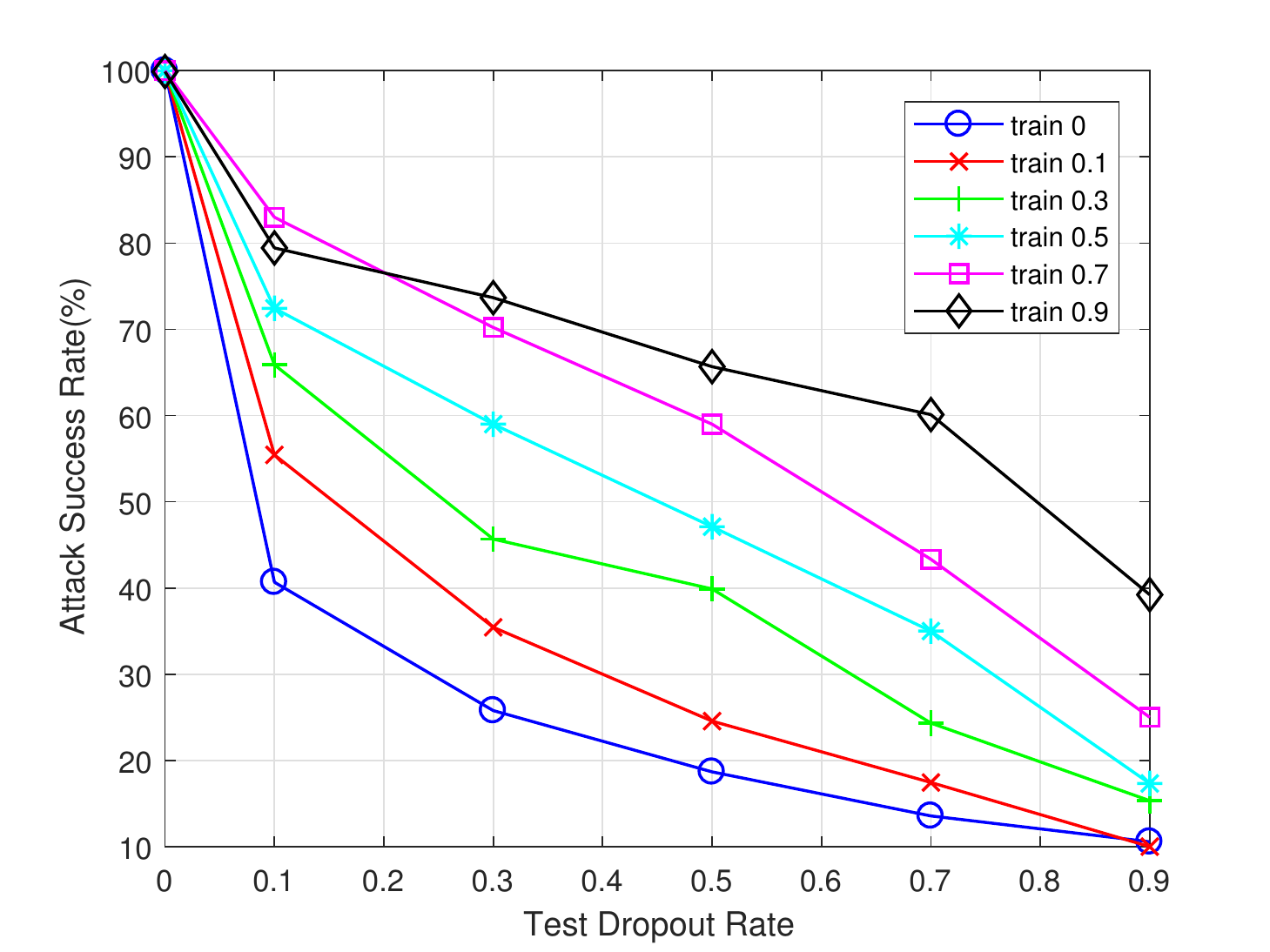}}
\subfigure[C\&W Attack $L_2$ Norm]{ 
\centering              
\includegraphics[width=1\textwidth]{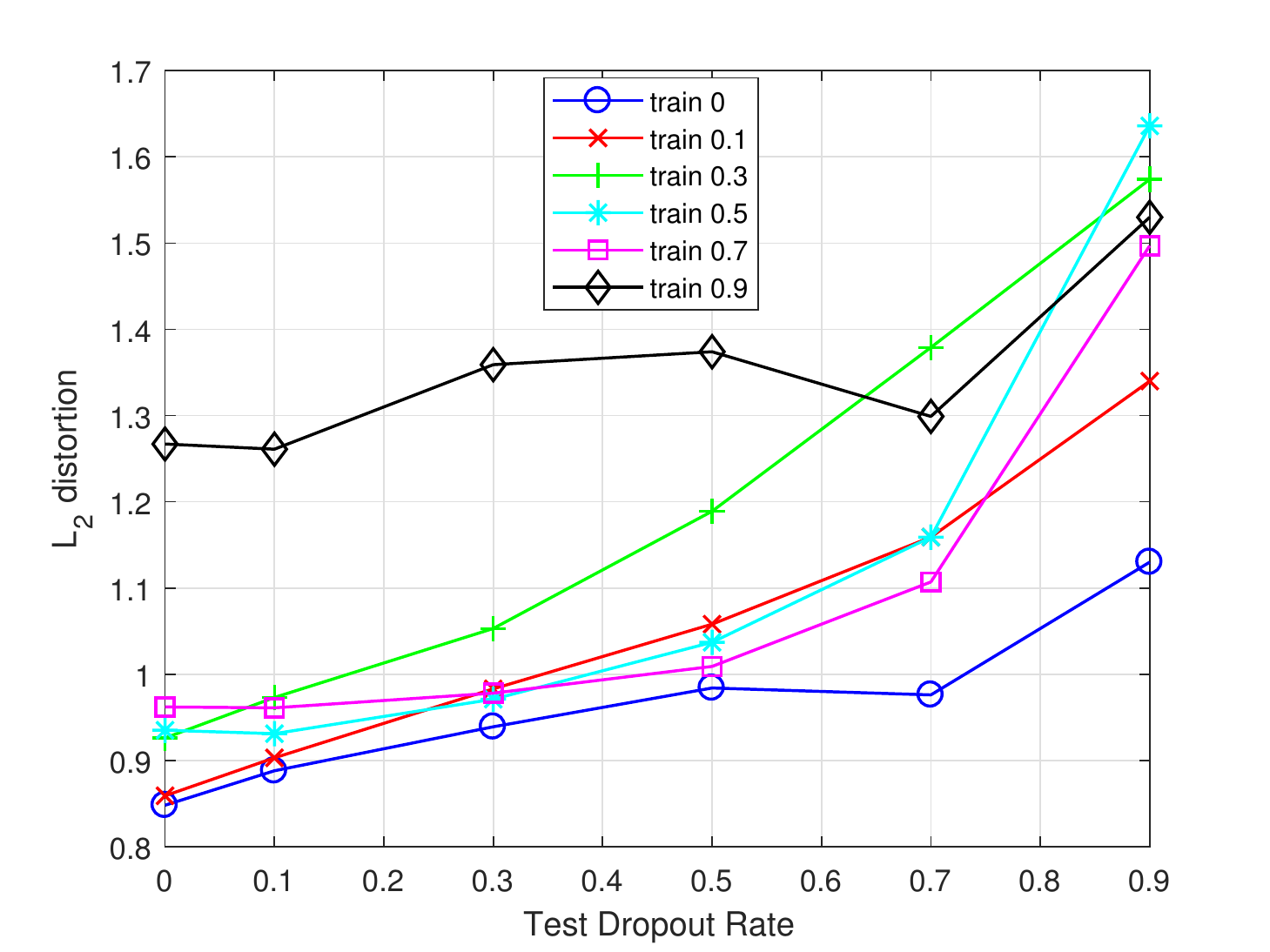}}
\end{minipage}
\caption{(a) Test accuracy, (b) Attack success rate, and (c) $L_2$ norm under C\&W attack on CIFAR-10 dataset using different training dropout rates and test dropout rates.}
\label{Fig:C&WCIFAR} 
\end{figure}

\setlength{\tabcolsep}{4pt}
\begin{table}
\begin{center}
\caption{Test accuracy, attack success rate, and $L_2$ norm using SAP against C\&W attack on CIFAR-10.}
\label{Table:SAPCIFAR}
\scalebox{0.88}{
\begin{tabular}{lllllll}
\hline\noalign{\smallskip}
  & train 0 & train 0.1 & train 0.3 & train 0.5 & train 0.7 & train 0.9 \\
\noalign{\smallskip}
\hline
\noalign{\smallskip}

Test acc.  & {\it} 72.07\% & 75.69\% & 76.39\% & 78.12\% & 78.15\% & 68.35\%\\
C\&W ASR  & {\it} 54.44\% & 64.44\% & 77.78\% & 78.89\% & 85.56\% & 70\%\\
$L_2$ norm & {\it} 0.504 & 0.522  & 0.618 & 0.498& 0.679 & 0.784\\
\hline
\end{tabular}}
\end{center}
\end{table}

\setlength{\tabcolsep}{1.4pt}

\setlength{\tabcolsep}{4pt}
\begin{table}
\begin{center}
\caption{Attack success rate using our defensive dropout against FGSM attack on CIFAR-10.}
\label{Fig:FGSMCIFAR}
\scalebox{0.9}{
\begin{tabular}{lllllll}
\hline\noalign{\smallskip}
Dropout rate & test 0 & test 0.1 & test 0.3 & test 0.5 & test 0.7 & test 0.9 \\
\noalign{\smallskip}
\hline
\noalign{\smallskip}
train 0 & {\it} 32.48\% & \quad -- & \quad -- & \quad -- &  \quad -- &  \quad --\\


train 0.5 & {\it} 15.87\% & 14.46\% & 13.89\% & \quad -- & \quad -- & \quad --\\
\hline
\end{tabular}}
\end{center}
\end{table}
\setlength{\tabcolsep}{1.4pt}

We first compare with SAP on the defense effects against C\&W attack using CIFAR-10 dataset.
The results of SAP are summarized in Table \ref{Table:SAPCIFAR}.
The results of our defensive dropout are summarized through Fig. \ref{Fig:C&WCIFAR}.
In Table \ref{Table:SAPCIFAR}, we perform SAP on neural network models using different training dropout rates (in columns).
The second to forth rows report test accuracy, attack success rate (ASR), and $L_2$ norm.
If we allow test accuracy decrease within 4\%, the SAP can reduce the attack success rate from 100\% to 77.78\% with a test accuracy of 76.39\%.
From Fig. \ref{Fig:C&WCIFAR}, we can observe that at training dropout rate of 0.7 and test dropout rate of 0.7, our defensive dropout can reduce the attack success rate to 43.33\% with a test accuracy of 77\%, demonstrating superior defense effects to SAP.
Also the $L_2$ norm of our defensive dropout is around 1.1 which is much higher than that of SAP (i.e., 0.618 from Table \ref{Table:SAPCIFAR}). 
Table \ref{Fig:FGSMCIFAR} shows the attack success rate using our defensive dropout against FGSM attack on CIFAR-10, observing that the defensive dropout reduces attack success rate from 32.48\% to 13.89\% at test accuracy of 77.89\%.
In general, attack success rate of FGSM is much lower than C\&W, because it is a faster, but not optimal attack.

We also summarize the results on MNIST dataset using our defensive dropout against FGSM, JSMA, and C\&W attacks, respectively, in Tables \ref{Fig:FGSMMNIST}, \ref{Fig:JSMAMNIST}, and \ref{Fig:CWMNIST},
with $<$1\% test accurary drop.
For FGSM, defensive dropout reduces attack success rate from 40.67\% to 16.44\%.
For JSMA, defensive dropout reduces attack success rate from 91.89\% to 26.78\%.
For C\&W, defensive dropout reduces attack success rate from 100\% to 13.89\%.
Please note that adversarial training and  distillation as a defense are totally vulnerable to C\&W attack \cite{carlini2017towards}. 

\setlength{\tabcolsep}{4pt}
\begin{table}
\begin{center}
\caption{Attack success rate using our defensive dropout against FGSM attack on MNIST.}
\label{Fig:FGSMMNIST}
\scalebox{0.9}{
\begin{tabular}{lllllll}
\hline\noalign{\smallskip}
Dropout rate & test 0 & test 0.1 & test 0.3 & test 0.5 & test 0.7 & test 0.9 \\
\noalign{\smallskip}
\hline
\noalign{\smallskip}


train 0.7 & {\it} 22.74\% & 21.89\% & 20.67\% & 19.56\% & 16.44\% & \quad --\\
\hline
\end{tabular}}
\end{center}
\end{table}
\setlength{\tabcolsep}{1.4pt}

\setlength{\tabcolsep}{4pt}
\begin{table}
\begin{center}
\caption{Attack success rate using our defensive dropout against JSMA attack on MNIST.}
\label{Fig:JSMAMNIST}
\scalebox{0.9}{
\begin{tabular}{lllllll}
\hline\noalign{\smallskip}
Dropout rate & test 0 & test 0.1 & test 0.3 & test 0.5 & test 0.7 & test 0.9 \\
\noalign{\smallskip}
\hline
\noalign{\smallskip}


train 0.7 & {\it} 90.67\% & 60.56\% & 43.78\% & 35.67\% & 26.78\% & \quad --\\
\hline
\end{tabular}}
\end{center}
\end{table}
\setlength{\tabcolsep}{1.4pt}

\setlength{\tabcolsep}{4pt}
\begin{table}
\begin{center}
\caption{Attack success rate using our defensive dropout against C\&W attack on MNIST.}
\label{Fig:CWMNIST}
\begin{tabular}{lllllll}
\hline\noalign{\smallskip}
Dropout rate & test 0 & test 0.1 & test 0.3 & test 0.5 & test 0.7 & test 0.9 \\
\noalign{\smallskip}
\hline
\noalign{\smallskip}


train 0.3 & {\it} 100\%& 24.66\% & 24.00\% & 13.89\% & \quad -- & \quad --\\
\hline
\end{tabular}
\end{center}
\end{table}
\setlength{\tabcolsep}{1.4pt}

\section{Conclusion}
In this paper, we propose defensive dropout for hardening deep neural networks under adversarial attacks. Considering the problem of building robust DNNs as an attacker-defender two-player game, we provide a defensive dropout algorithm that determines an optimal test dropout rate given the neural network model and the attacker's strategy for generating adversarial examples. We also explain the mechanism behind the outstanding defense effects achieved by the proposed defensive dropout. 

\section*{Acknowledgements}
This work is supported by the National Science Foundation (CCF-1733701, CNS-1618379, DMS-1737897, and CNS-1840813), Air Force Research Laboratory FA8750-18-2-0058, Naval Research Laboratory. We thank researchers at the US Naval Research Laboratory for their comments on previous drafts of this paper.


\bibliographystyle{ACM-Reference-Format}
\bibliography{sample-bibliography,egbib}


\begin{thebibliography}{34}


\ifx \showCODEN    \undefined \def \showCODEN     #1{\unskip}     \fi
\ifx \showDOI      \undefined \def \showDOI       #1{#1}\fi
\ifx \showISBNx    \undefined \def \showISBNx     #1{\unskip}     \fi
\ifx \showISBNxiii \undefined \def \showISBNxiii  #1{\unskip}     \fi
\ifx \showISSN     \undefined \def \showISSN      #1{\unskip}     \fi
\ifx \showLCCN     \undefined \def \showLCCN      #1{\unskip}     \fi
\ifx \shownote     \undefined \def \shownote      #1{#1}          \fi
\ifx \showarticletitle \undefined \def \showarticletitle #1{#1}   \fi
\ifx \showURL      \undefined \def \showURL       {\relax}        \fi
\providecommand\bibfield[2]{#2}
\providecommand\bibinfo[2]{#2}
\providecommand\natexlab[1]{#1}
\providecommand\showeprint[2][]{arXiv:#2}

\bibitem[\protect\citeauthoryear{Athalye, Carlini, and Wagner}{Athalye
  et~al\mbox{.}}{2018}]%
        {athalye2018obfuscated}
\bibfield{author}{\bibinfo{person}{Anish Athalye}, \bibinfo{person}{Nicholas
  Carlini}, {and} \bibinfo{person}{David Wagner}.}
  \bibinfo{year}{2018}\natexlab{}.
\newblock \showarticletitle{Obfuscated gradients give a false sense of
  security: Circumventing defenses to adversarial examples}.
\newblock \bibinfo{journal}{\emph{arXiv preprint arXiv:1802.00420}}
  (\bibinfo{year}{2018}).
\newblock


\bibitem[\protect\citeauthoryear{Bhagoji, Cullina, and Mittal}{Bhagoji
  et~al\mbox{.}}{2017}]%
        {bhagoji2017dimensionality}
\bibfield{author}{\bibinfo{person}{Arjun~Nitin Bhagoji},
  \bibinfo{person}{Daniel Cullina}, {and} \bibinfo{person}{Prateek Mittal}.}
  \bibinfo{year}{2017}\natexlab{}.
\newblock \showarticletitle{Dimensionality reduction as a defense against
  evasion attacks on machine learning classifiers}.
\newblock \bibinfo{journal}{\emph{arXiv preprint arXiv:1704.02654}}
  (\bibinfo{year}{2017}).
\newblock


\bibitem[\protect\citeauthoryear{Bouthillier, Konda, Vincent, and
  Memisevic}{Bouthillier et~al\mbox{.}}{2015}]%
        {bouthillier2015dropout}
\bibfield{author}{\bibinfo{person}{Xavier Bouthillier},
  \bibinfo{person}{Kishore Konda}, \bibinfo{person}{Pascal Vincent}, {and}
  \bibinfo{person}{Roland Memisevic}.} \bibinfo{year}{2015}\natexlab{}.
\newblock \showarticletitle{Dropout as data augmentation}.
\newblock \bibinfo{journal}{\emph{arXiv preprint arXiv:1506.08700}}
  (\bibinfo{year}{2015}).
\newblock


\bibitem[\protect\citeauthoryear{Carlini, Mishra, Vaidya, Zhang, Sherr,
  Shields, Wagner, and Zhou}{Carlini et~al\mbox{.}}{2016}]%
        {carlini2016hidden}
\bibfield{author}{\bibinfo{person}{Nicholas Carlini}, \bibinfo{person}{Pratyush
  Mishra}, \bibinfo{person}{Tavish Vaidya}, \bibinfo{person}{Yuankai Zhang},
  \bibinfo{person}{Micah Sherr}, \bibinfo{person}{Clay Shields},
  \bibinfo{person}{David Wagner}, {and} \bibinfo{person}{Wenchao Zhou}.}
  \bibinfo{year}{2016}\natexlab{}.
\newblock \showarticletitle{Hidden Voice Commands.}. In
  \bibinfo{booktitle}{\emph{USENIX Security Symposium}}.
  \bibinfo{pages}{513--530}.
\newblock


\bibitem[\protect\citeauthoryear{Carlini and Wagner}{Carlini and
  Wagner}{2017}]%
        {carlini2017towards}
\bibfield{author}{\bibinfo{person}{Nicholas Carlini} {and}
  \bibinfo{person}{David Wagner}.} \bibinfo{year}{2017}\natexlab{}.
\newblock \showarticletitle{Towards evaluating the robustness of neural
  networks}. In \bibinfo{booktitle}{\emph{Security and Privacy (SP), 2017 IEEE
  Symposium on}}. IEEE, \bibinfo{pages}{39--57}.
\newblock


\bibitem[\protect\citeauthoryear{Chen, Sharma, Zhang, Yi, and Hsieh}{Chen
  et~al\mbox{.}}{2017}]%
        {chen2017ead}
\bibfield{author}{\bibinfo{person}{Pin-Yu Chen}, \bibinfo{person}{Yash Sharma},
  \bibinfo{person}{Huan Zhang}, \bibinfo{person}{Jinfeng Yi}, {and}
  \bibinfo{person}{Cho-Jui Hsieh}.} \bibinfo{year}{2017}\natexlab{}.
\newblock \showarticletitle{EAD: elastic-net attacks to deep neural networks
  via adversarial examples}.
\newblock \bibinfo{journal}{\emph{arXiv preprint arXiv:1709.04114}}
  (\bibinfo{year}{2017}).
\newblock


\bibitem[\protect\citeauthoryear{Collobert and Weston}{Collobert and
  Weston}{2008}]%
        {collobert2008unified}
\bibfield{author}{\bibinfo{person}{Ronan Collobert} {and}
  \bibinfo{person}{Jason Weston}.} \bibinfo{year}{2008}\natexlab{}.
\newblock \showarticletitle{A unified architecture for natural language
  processing: Deep neural networks with multitask learning}. In
  \bibinfo{booktitle}{\emph{Proceedings of the 25th international conference on
  Machine learning}}. ACM, \bibinfo{pages}{160--167}.
\newblock


\bibitem[\protect\citeauthoryear{Dhillon, Azizzadenesheli, Bernstein, Kossaifi,
  Khanna, Lipton, and Anandkumar}{Dhillon et~al\mbox{.}}{2018}]%
        {s.2018stochastic}
\bibfield{author}{\bibinfo{person}{Guneet~S. Dhillon}, \bibinfo{person}{Kamyar
  Azizzadenesheli}, \bibinfo{person}{Jeremy~D. Bernstein},
  \bibinfo{person}{Jean Kossaifi}, \bibinfo{person}{Aran Khanna},
  \bibinfo{person}{Zachary~C. Lipton}, {and} \bibinfo{person}{Animashree
  Anandkumar}.} \bibinfo{year}{2018}\natexlab{}.
\newblock \showarticletitle{Stochastic activation pruning for robust
  adversarial defense}. In \bibinfo{booktitle}{\emph{International Conference
  on Learning Representations}}.
\newblock
\urldef\tempurl%
\url{https://openreview.net/forum?id=H1uR4GZRZ}
\showURL{%
\tempurl}


\bibitem[\protect\citeauthoryear{Dziugaite, Ghahramani, and Roy}{Dziugaite
  et~al\mbox{.}}{2016}]%
        {dziugaite2016study}
\bibfield{author}{\bibinfo{person}{Gintare~Karolina Dziugaite},
  \bibinfo{person}{Zoubin Ghahramani}, {and} \bibinfo{person}{Daniel~M Roy}.}
  \bibinfo{year}{2016}\natexlab{}.
\newblock \showarticletitle{A study of the effect of jpg compression on
  adversarial images}.
\newblock \bibinfo{journal}{\emph{arXiv preprint arXiv:1608.00853}}
  (\bibinfo{year}{2016}).
\newblock


\bibitem[\protect\citeauthoryear{Feinman, Curtin, Shintre, and Gardner}{Feinman
  et~al\mbox{.}}{2017}]%
        {feinman2017detecting}
\bibfield{author}{\bibinfo{person}{Reuben Feinman}, \bibinfo{person}{Ryan~R
  Curtin}, \bibinfo{person}{Saurabh Shintre}, {and} \bibinfo{person}{Andrew~B
  Gardner}.} \bibinfo{year}{2017}\natexlab{}.
\newblock \showarticletitle{Detecting adversarial samples from artifacts}.
\newblock \bibinfo{journal}{\emph{arXiv preprint arXiv:1703.00410}}
  (\bibinfo{year}{2017}).
\newblock


\bibitem[\protect\citeauthoryear{Goodfellow, Bengio, Courville, and
  Bengio}{Goodfellow et~al\mbox{.}}{2016}]%
        {goodfellow2016deep}
\bibfield{author}{\bibinfo{person}{Ian Goodfellow}, \bibinfo{person}{Yoshua
  Bengio}, \bibinfo{person}{Aaron Courville}, {and} \bibinfo{person}{Yoshua
  Bengio}.} \bibinfo{year}{2016}\natexlab{}.
\newblock \bibinfo{booktitle}{\emph{Deep learning}}. Vol.~\bibinfo{volume}{1}.
\newblock \bibinfo{publisher}{MIT press Cambridge}.
\newblock


\bibitem[\protect\citeauthoryear{Goodfellow, Shlens, and Szegedy}{Goodfellow
  et~al\mbox{.}}{2014}]%
        {goodfellow2014explaining}
\bibfield{author}{\bibinfo{person}{Ian~J Goodfellow}, \bibinfo{person}{Jonathon
  Shlens}, {and} \bibinfo{person}{Christian Szegedy}.}
  \bibinfo{year}{2014}\natexlab{}.
\newblock \showarticletitle{Explaining and harnessing adversarial examples}.
\newblock \bibinfo{journal}{\emph{arXiv preprint arXiv:1412.6572}}
  (\bibinfo{year}{2014}).
\newblock


\bibitem[\protect\citeauthoryear{Guo, Rana, Ciss{\'e}, and van~der Maaten}{Guo
  et~al\mbox{.}}{2017}]%
        {guo2017countering}
\bibfield{author}{\bibinfo{person}{Chuan Guo}, \bibinfo{person}{Mayank Rana},
  \bibinfo{person}{Moustapha Ciss{\'e}}, {and} \bibinfo{person}{Laurens van~der
  Maaten}.} \bibinfo{year}{2017}\natexlab{}.
\newblock \showarticletitle{Countering Adversarial Images using Input
  Transformations}.
\newblock \bibinfo{journal}{\emph{arXiv preprint arXiv:1711.00117}}
  (\bibinfo{year}{2017}).
\newblock


\bibitem[\protect\citeauthoryear{Hinton, Deng, Yu, Dahl, Mohamed, Jaitly,
  Senior, Vanhoucke, Nguyen, Sainath, et~al\mbox{.}}{Hinton
  et~al\mbox{.}}{2012a}]%
        {hinton2012deep}
\bibfield{author}{\bibinfo{person}{Geoffrey Hinton}, \bibinfo{person}{Li Deng},
  \bibinfo{person}{Dong Yu}, \bibinfo{person}{George~E Dahl},
  \bibinfo{person}{Abdel-rahman Mohamed}, \bibinfo{person}{Navdeep Jaitly},
  \bibinfo{person}{Andrew Senior}, \bibinfo{person}{Vincent Vanhoucke},
  \bibinfo{person}{Patrick Nguyen}, \bibinfo{person}{Tara~N Sainath},
  {et~al\mbox{.}}} \bibinfo{year}{2012}\natexlab{a}.
\newblock \showarticletitle{Deep neural networks for acoustic modeling in
  speech recognition: The shared views of four research groups}.
\newblock \bibinfo{journal}{\emph{IEEE Signal Processing Magazine}}
  \bibinfo{volume}{29}, \bibinfo{number}{6} (\bibinfo{year}{2012}),
  \bibinfo{pages}{82--97}.
\newblock


\bibitem[\protect\citeauthoryear{Hinton, Srivastava, Krizhevsky, Sutskever, and
  Salakhutdinov}{Hinton et~al\mbox{.}}{2012b}]%
        {hinton2012improving}
\bibfield{author}{\bibinfo{person}{Geoffrey~E Hinton}, \bibinfo{person}{Nitish
  Srivastava}, \bibinfo{person}{Alex Krizhevsky}, \bibinfo{person}{Ilya
  Sutskever}, {and} \bibinfo{person}{Ruslan~R Salakhutdinov}.}
  \bibinfo{year}{2012}\natexlab{b}.
\newblock \showarticletitle{Improving neural networks by preventing
  co-adaptation of feature detectors}.
\newblock \bibinfo{journal}{\emph{arXiv preprint arXiv:1207.0580}}
  (\bibinfo{year}{2012}).
\newblock


\bibitem[\protect\citeauthoryear{Karpathy and Fei-Fei}{Karpathy and
  Fei-Fei}{2015}]%
        {karpathy2015deep}
\bibfield{author}{\bibinfo{person}{Andrej Karpathy} {and} \bibinfo{person}{Li
  Fei-Fei}.} \bibinfo{year}{2015}\natexlab{}.
\newblock \showarticletitle{Deep visual-semantic alignments for generating
  image descriptions}. In \bibinfo{booktitle}{\emph{Proceedings of the IEEE
  conference on computer vision and pattern recognition}}.
  \bibinfo{pages}{3128--3137}.
\newblock


\bibitem[\protect\citeauthoryear{Kingma and Ba}{Kingma and Ba}{2015}]%
        {KingmaB2015adam}
\bibfield{author}{\bibinfo{person}{Diederik~P. Kingma} {and}
  \bibinfo{person}{Jimmy Ba}.} \bibinfo{year}{2015}\natexlab{}.
\newblock \showarticletitle{Adam: {A} Method for Stochastic Optimization}.
\newblock \bibinfo{journal}{\emph{2015 ICLR}}  \bibinfo{volume}{arXiv preprint
  arXiv:1412.6980} (\bibinfo{year}{2015}).
\newblock
\showeprint[arxiv]{1412.6980}
\urldef\tempurl%
\url{http://arxiv.org/abs/1412.6980}
\showURL{%
\tempurl}


\bibitem[\protect\citeauthoryear{Krizhevsky and Hinton}{Krizhevsky and
  Hinton}{2009}]%
        {krizhevsky2009learning}
\bibfield{author}{\bibinfo{person}{Alex Krizhevsky} {and}
  \bibinfo{person}{Geoffrey Hinton}.} \bibinfo{year}{2009}\natexlab{}.
\newblock \showarticletitle{Learning multiple layers of features from tiny
  images}.
\newblock  (\bibinfo{year}{2009}).
\newblock


\bibitem[\protect\citeauthoryear{Krizhevsky, Sutskever, and Hinton}{Krizhevsky
  et~al\mbox{.}}{2012}]%
        {krizhevsky2012imagenet}
\bibfield{author}{\bibinfo{person}{Alex Krizhevsky}, \bibinfo{person}{Ilya
  Sutskever}, {and} \bibinfo{person}{Geoffrey~E Hinton}.}
  \bibinfo{year}{2012}\natexlab{}.
\newblock \showarticletitle{Imagenet classification with deep convolutional
  neural networks}. In \bibinfo{booktitle}{\emph{Advances in neural information
  processing systems}}. \bibinfo{pages}{1097--1105}.
\newblock


\bibitem[\protect\citeauthoryear{Kurakin, Goodfellow, and Bengio}{Kurakin
  et~al\mbox{.}}{2016}]%
        {kurakin2016adversarial}
\bibfield{author}{\bibinfo{person}{Alexey Kurakin}, \bibinfo{person}{Ian
  Goodfellow}, {and} \bibinfo{person}{Samy Bengio}.}
  \bibinfo{year}{2016}\natexlab{}.
\newblock \showarticletitle{Adversarial examples in the physical world}.
\newblock \bibinfo{journal}{\emph{arXiv preprint arXiv:1607.02533}}
  (\bibinfo{year}{2016}).
\newblock


\bibitem[\protect\citeauthoryear{LeCun, Boser, Denker, Henderson, Howard,
  Hubbard, and Jackel}{LeCun et~al\mbox{.}}{1989}]%
        {lecun1989backpropagation}
\bibfield{author}{\bibinfo{person}{Yann LeCun}, \bibinfo{person}{Bernhard
  Boser}, \bibinfo{person}{John~S Denker}, \bibinfo{person}{Donnie Henderson},
  \bibinfo{person}{Richard~E Howard}, \bibinfo{person}{Wayne Hubbard}, {and}
  \bibinfo{person}{Lawrence~D Jackel}.} \bibinfo{year}{1989}\natexlab{}.
\newblock \showarticletitle{Backpropagation applied to handwritten zip code
  recognition}.
\newblock \bibinfo{journal}{\emph{Neural computation}} \bibinfo{volume}{1},
  \bibinfo{number}{4} (\bibinfo{year}{1989}), \bibinfo{pages}{541--551}.
\newblock


\bibitem[\protect\citeauthoryear{LeCun, Bottou, Bengio, and Haffner}{LeCun
  et~al\mbox{.}}{1998}]%
        {lecun1998gradient}
\bibfield{author}{\bibinfo{person}{Yann LeCun}, \bibinfo{person}{L{\'e}on
  Bottou}, \bibinfo{person}{Yoshua Bengio}, {and} \bibinfo{person}{Patrick
  Haffner}.} \bibinfo{year}{1998}\natexlab{}.
\newblock \showarticletitle{Gradient-based learning applied to document
  recognition}.
\newblock \bibinfo{journal}{\emph{Proc. IEEE}} \bibinfo{volume}{86},
  \bibinfo{number}{11} (\bibinfo{year}{1998}), \bibinfo{pages}{2278--2324}.
\newblock


\bibitem[\protect\citeauthoryear{Liu, Wei, Luo, and Xu}{Liu
  et~al\mbox{.}}{2017}]%
        {liu2017fault}
\bibfield{author}{\bibinfo{person}{Yannan Liu}, \bibinfo{person}{Lingxiao Wei},
  \bibinfo{person}{Bo Luo}, {and} \bibinfo{person}{Qiang Xu}.}
  \bibinfo{year}{2017}\natexlab{}.
\newblock \showarticletitle{Fault injection attack on deep neural network}. In
  \bibinfo{booktitle}{\emph{Proceedings of the 36th International Conference on
  Computer-Aided Design}}. IEEE Press, \bibinfo{pages}{131--138}.
\newblock


\bibitem[\protect\citeauthoryear{Livnat, Papadimitriou, Pippenger, and
  Feldman}{Livnat et~al\mbox{.}}{2010}]%
        {livnat2010sex}
\bibfield{author}{\bibinfo{person}{Adi Livnat}, \bibinfo{person}{Christos
  Papadimitriou}, \bibinfo{person}{Nicholas Pippenger}, {and}
  \bibinfo{person}{Marcus~W Feldman}.} \bibinfo{year}{2010}\natexlab{}.
\newblock \showarticletitle{Sex, mixability, and modularity}.
\newblock \bibinfo{journal}{\emph{Proceedings of the National Academy of
  Sciences}} \bibinfo{volume}{107}, \bibinfo{number}{4} (\bibinfo{year}{2010}),
  \bibinfo{pages}{1452--1457}.
\newblock


\bibitem[\protect\citeauthoryear{Nguyen, Yosinski, and Clune}{Nguyen
  et~al\mbox{.}}{2015}]%
        {nguyen2015deep}
\bibfield{author}{\bibinfo{person}{Anh Nguyen}, \bibinfo{person}{Jason
  Yosinski}, {and} \bibinfo{person}{Jeff Clune}.}
  \bibinfo{year}{2015}\natexlab{}.
\newblock \showarticletitle{Deep neural networks are easily fooled: High
  confidence predictions for unrecognizable images}. In
  \bibinfo{booktitle}{\emph{Proceedings of the IEEE Conference on Computer
  Vision and Pattern Recognition}}. \bibinfo{pages}{427--436}.
\newblock


\bibitem[\protect\citeauthoryear{Papernot, Carlini, Goodfellow, Feinman,
  Faghri, Matyasko, Hambardzumyan, Juang, Kurakin, Sheatsley,
  et~al\mbox{.}}{Papernot et~al\mbox{.}}{2016a}]%
        {papernot2016cleverhans}
\bibfield{author}{\bibinfo{person}{Nicolas Papernot}, \bibinfo{person}{Nicholas
  Carlini}, \bibinfo{person}{Ian Goodfellow}, \bibinfo{person}{Reuben Feinman},
  \bibinfo{person}{Fartash Faghri}, \bibinfo{person}{Alexander Matyasko},
  \bibinfo{person}{Karen Hambardzumyan}, \bibinfo{person}{Yi-Lin Juang},
  \bibinfo{person}{Alexey Kurakin}, \bibinfo{person}{Ryan Sheatsley},
  {et~al\mbox{.}}} \bibinfo{year}{2016}\natexlab{a}.
\newblock \showarticletitle{cleverhans v2. 0.0: an adversarial machine learning
  library}.
\newblock \bibinfo{journal}{\emph{arXiv preprint arXiv:1610.00768}}
  (\bibinfo{year}{2016}).
\newblock


\bibitem[\protect\citeauthoryear{Papernot, McDaniel, Jha, Fredrikson, Celik,
  and Swami}{Papernot et~al\mbox{.}}{2016b}]%
        {papernot2016limitations}
\bibfield{author}{\bibinfo{person}{Nicolas Papernot}, \bibinfo{person}{Patrick
  McDaniel}, \bibinfo{person}{Somesh Jha}, \bibinfo{person}{Matt Fredrikson},
  \bibinfo{person}{Z~Berkay Celik}, {and} \bibinfo{person}{Ananthram Swami}.}
  \bibinfo{year}{2016}\natexlab{b}.
\newblock \showarticletitle{The limitations of deep learning in adversarial
  settings}. In \bibinfo{booktitle}{\emph{Security and Privacy (EuroS\&P), 2016
  IEEE European Symposium on}}. IEEE, \bibinfo{pages}{372--387}.
\newblock


\bibitem[\protect\citeauthoryear{Papernot, McDaniel, Wu, Jha, and
  Swami}{Papernot et~al\mbox{.}}{2016c}]%
        {papernot2016distillation}
\bibfield{author}{\bibinfo{person}{Nicolas Papernot}, \bibinfo{person}{Patrick
  McDaniel}, \bibinfo{person}{Xi Wu}, \bibinfo{person}{Somesh Jha}, {and}
  \bibinfo{person}{Ananthram Swami}.} \bibinfo{year}{2016}\natexlab{c}.
\newblock \showarticletitle{Distillation as a defense to adversarial
  perturbations against deep neural networks}. In
  \bibinfo{booktitle}{\emph{Security and Privacy (SP), 2016 IEEE Symposium
  on}}. IEEE, \bibinfo{pages}{582--597}.
\newblock


\bibitem[\protect\citeauthoryear{Srivastava, Hinton, Krizhevsky, Sutskever, and
  Salakhutdinov}{Srivastava et~al\mbox{.}}{2014}]%
        {srivastava2014dropout}
\bibfield{author}{\bibinfo{person}{Nitish Srivastava},
  \bibinfo{person}{Geoffrey Hinton}, \bibinfo{person}{Alex Krizhevsky},
  \bibinfo{person}{Ilya Sutskever}, {and} \bibinfo{person}{Ruslan
  Salakhutdinov}.} \bibinfo{year}{2014}\natexlab{}.
\newblock \showarticletitle{Dropout: A simple way to prevent neural networks
  from overfitting}.
\newblock \bibinfo{journal}{\emph{The Journal of Machine Learning Research}}
  \bibinfo{volume}{15}, \bibinfo{number}{1} (\bibinfo{year}{2014}),
  \bibinfo{pages}{1929--1958}.
\newblock


\bibitem[\protect\citeauthoryear{Szegedy, Zaremba, Sutskever, Bruna, Erhan,
  Goodfellow, and Fergus}{Szegedy et~al\mbox{.}}{2013}]%
        {szegedy2013intriguing}
\bibfield{author}{\bibinfo{person}{Christian Szegedy},
  \bibinfo{person}{Wojciech Zaremba}, \bibinfo{person}{Ilya Sutskever},
  \bibinfo{person}{Joan Bruna}, \bibinfo{person}{Dumitru Erhan},
  \bibinfo{person}{Ian Goodfellow}, {and} \bibinfo{person}{Rob Fergus}.}
  \bibinfo{year}{2013}\natexlab{}.
\newblock \showarticletitle{Intriguing properties of neural networks}.
\newblock \bibinfo{journal}{\emph{arXiv preprint arXiv:1312.6199}}
  (\bibinfo{year}{2013}).
\newblock


\bibitem[\protect\citeauthoryear{Tram{\`e}r, Kurakin, Papernot, Boneh, and
  McDaniel}{Tram{\`e}r et~al\mbox{.}}{2017}]%
        {tramer2017ensemble}
\bibfield{author}{\bibinfo{person}{Florian Tram{\`e}r}, \bibinfo{person}{Alexey
  Kurakin}, \bibinfo{person}{Nicolas Papernot}, \bibinfo{person}{Dan Boneh},
  {and} \bibinfo{person}{Patrick McDaniel}.} \bibinfo{year}{2017}\natexlab{}.
\newblock \showarticletitle{Ensemble adversarial training: Attacks and
  defenses}.
\newblock \bibinfo{journal}{\emph{arXiv preprint arXiv:1705.07204}}
  (\bibinfo{year}{2017}).
\newblock


\bibitem[\protect\citeauthoryear{{Tram{\`e}r}, {Kurakin}, {Papernot},
  {Goodfellow}, {Boneh}, and {McDaniel}}{{Tram{\`e}r} et~al\mbox{.}}{2018}]%
        {tram2018ensemble}
\bibfield{author}{\bibinfo{person}{F. {Tram{\`e}r}}, \bibinfo{person}{A.
  {Kurakin}}, \bibinfo{person}{N. {Papernot}}, \bibinfo{person}{I.
  {Goodfellow}}, \bibinfo{person}{D. {Boneh}}, {and} \bibinfo{person}{P.
  {McDaniel}}.} \bibinfo{year}{2018}\natexlab{}.
\newblock \showarticletitle{Ensemble Adversarial Training: Attacks and
  Defenses}.
\newblock \bibinfo{journal}{\emph{2018 ICLR}}  \bibinfo{volume}{arXiv preprint
  arXiv:1705.07204} (\bibinfo{year}{2018}).
\newblock


\bibitem[\protect\citeauthoryear{Xie, Wang, Zhang, Ren, and Yuille}{Xie
  et~al\mbox{.}}{2017}]%
        {xie2017mitigating}
\bibfield{author}{\bibinfo{person}{Cihang Xie}, \bibinfo{person}{Jianyu Wang},
  \bibinfo{person}{Zhishuai Zhang}, \bibinfo{person}{Zhou Ren}, {and}
  \bibinfo{person}{Alan Yuille}.} \bibinfo{year}{2017}\natexlab{}.
\newblock \showarticletitle{Mitigating adversarial effects through
  randomization}.
\newblock \bibinfo{journal}{\emph{arXiv preprint arXiv:1711.01991}}
  (\bibinfo{year}{2017}).
\newblock


\bibitem[\protect\citeauthoryear{Yann, Corinna, and Christopher}{Yann
  et~al\mbox{.}}{1998}]%
        {yann1998mnist}
\bibfield{author}{\bibinfo{person}{LeCun Yann}, \bibinfo{person}{Cortes
  Corinna}, {and} \bibinfo{person}{JB Christopher}.}
  \bibinfo{year}{1998}\natexlab{}.
\newblock \showarticletitle{The MNIST database of handwritten digits}.
\newblock \bibinfo{journal}{\emph{URL http://yhann. lecun. com/exdb/mnist}}
  (\bibinfo{year}{1998}).
\newblock


\end{thebibliography}

\end{document}